# Rotation-managed dissipative solitons

Yaroslav V. Kartashov[1,2,*] and Lluis Torner[1]

[1]ICFO-Institut de Ciencies Fotoniques and Universitat Politecnica de Catalunya, 08860 Castelldefels (Barcelona), Spain
[2]Institute of Spectroscopy, Russian Academy of Sciences, Troitsk, Moscow Region, 142190, Russia
*Corresponding author: Yaroslav.Kartashov@icfo.eu

We show that when spatially localized gain landscape performs accelerated motion in the transverse plane, i.e., when it rotates or oscillates around the light propagation axis, the effective gain experienced by the light beam considerably reduces with increase of the amplitude of oscillations or frequency of rotation of the localized gain. In the presence of uniform background losses and defocusing nonlinearity, such gain landscapes may support dynamically oscillating gain-managed solitons, but if the amplitude of oscillations or the frequency of rotation of the localized gain exceeds a threshold, stable attractors disappear and any input beam decays.

Dissipative optical systems afford the possibility to explore the interplay of a variety of different physical effects that act on light beams, since in addition to the well-known conservative effects such as diffraction, dispersion, and self-action, light beams propagating in such media experience gain and losses, which sometimes may have different physical origins, or may be spatially inhomogeneous. If a simultaneous stable balance is possible between gain and losses acting in the medium, and between nonlinearity and diffraction/dispersion, formation of stable dissipative solitons – which thus appear as attractors of the system - becomes possible [1,2]. Such dissipative solitons have been studied in a variety of optical systems governed by the Ginzburg-Landau equation [3-5] in lasers with saturable gain and absorption [2], semiconductor amplifiers [6], and in media with spatially inhomogeneous gain [7-11] or absorption [12].

Formation of solitons in media with spatially inhomogeneous gain is closely related to the effect of gain-guiding that enables stationary beam evolution for specific gain/losses profiles even in linear media and in the absence of conventional waveguides [13]. Recently stable solitons bifurcating from gain-guided modes were predicted in defocusing cubic media [14]. Similar effects can be encountered in other physical systems, such as non-conservative polariton condensates, whose evolution is governed by a similar mathematical model [15]. Up to now, only soliton formation in static gain profiles has been studied. Thus, a question arises about whether a transverse dynamics of the gain landscape may lead to qualitative modifications in the evolution of light beams.

In this Letter we show that the accelerated motion of the amplifying channel in the transverse plane results in the modification of the entire effective gain profile experienced by the propagating beam and an increase of the gain strength for which gain-guiding is possible. In the presence of background losses and defocusing nonlinearity, oscillating or rotating gain landscapes can support gain-managed solitons, whose parameters are dictated by the amplitude of oscillations or the frequency of rotation of the gain profile.

We describe the propagation of a light beam along the $\xi$-axis of the defocusing Kerr medium with a spatially inhomogeneous gain profile by the nonlinear Schrödinger equation for the dimensionless light field amplitude $q$:

$$i\frac{\partial q}{\partial \xi} = -\frac{1}{2}\Delta_\perp q + q|q|^2 + i(p_\text{i} R - \gamma)q \qquad (1)$$

Here $\xi$ is the propagation distance scaled to the diffraction length, $\Delta_\perp = \partial^2/\partial\eta^2 + \partial^2/\partial\zeta^2$ is the Laplacian (for the one-dimensional case it reads $\Delta_\perp = \partial^2/\partial\eta^2$), $\eta, \zeta$ are the transverse coordinates scaled to the characteristic width, the parameter $\gamma$ describes uniform losses, while $p_\text{i} > \gamma$ is the gain strength, and the function $R(\eta,\zeta,\xi)$ describes the dynamical gain profile. Dynamical gain landscapes can be created in materials with a spatially-inhomogeneous concentration of active dopants or by using suitably shaped (for example, bending along the $\xi$ axis) pumping beams. Further, we set $\gamma = 1$ and consider gain landscapes with $\max[R(\eta,\zeta,\xi)] = 1$.

We start from the one-dimensional case and consider a Gaussian gain landscape $R(\eta,\xi) = \exp[-(\eta - a\sin(\Omega\xi))^2]$ that oscillates in the transverse plane with the amplitude $a$ and frequency $\Omega$. It was shown [14] that at $a=0$ the static gain profile can support linear gain-guided modes for the unique gain strength $p_\text{i} = p_\text{i}^\text{cr}$ (for our set of parameters $p_\text{i}^\text{cr} \approx 1.63$). The addition of nonlinearity allows us to obtain solitons in the domains $p_\text{i} < p_\text{i}^\text{cr}$ (in focusing medium) and $p_\text{i} > p_\text{i}^\text{cr}$ (in defocusing medium). Static states are stable only in the defocusing medium.

Figure 1 shows the evolution of the beam in the oscillating gain landscape. Instead of static solitons, after a sufficiently long propagation distance [the initial stage of evolution is not shown in Fig. 1(c)], the dynamically oscillating "gain-managed" state is formed that exactly reproduces its shape after each period of oscillations of the gain channel. Physically, such solitons form because defocusing nonlinearity expels light from the central amplifying domain to the periphery, where power is absorbed. It is important to note that, despite considerable radiation from the amplifying domain, the soliton has exponentially decaying tails at any distance. The power $U = \int_{-\infty}^{\infty} |q|^2\, d\eta$ slightly oscillates around a certain averaged value with a period $\pi/\Omega$ (recall that the soliton profile is reproduced

after $\xi=2\pi/\Omega$). The field modulus distributions at the distances $\xi$ and $\xi+\pi/\Omega$ are mirror images of each other [Figs. 1(a) and 1(b)]. These figures also show internal currents $j=u^2\partial\varphi/\partial\eta$, where $u=|q|$ and $\varphi=\arg(q)$, indicating that there is always power flow from the amplifying domain toward the periphery.

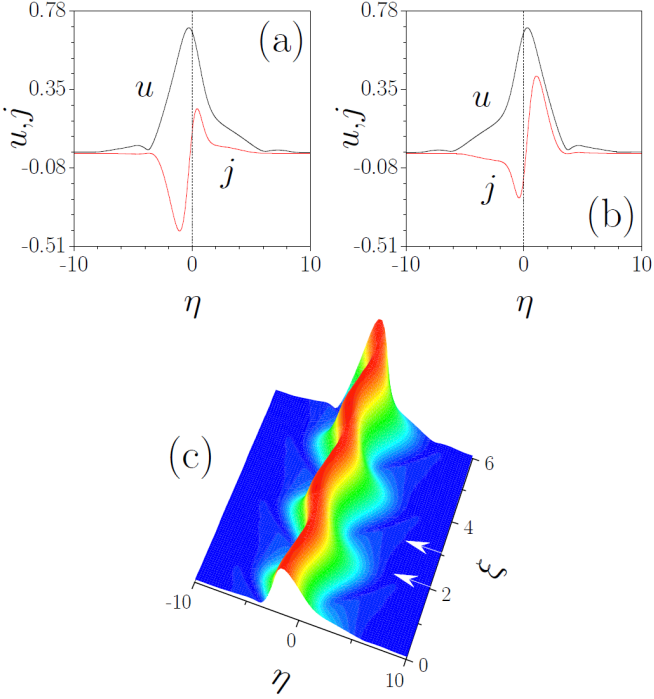

Fig. 1. (a),(b) Instantaneous field modulus distributions and current profiles in a dynamically oscillating soliton corresponding to the points marked by arrows in panel (c) that shows propagation dynamics at $a=0.5$, $p_i=1.8$, $\Omega=\pi$. The current in (a),(b) is divided by 50 for illustrative purposes.

Figure 1 implies that the soliton can be represented as a sum of a "guiding center" [16] and oscillating parts. At $a\sin(\Omega\xi)=\varepsilon\ll1$ the expansion of the gain profile into series in $\varepsilon$ up to $\varepsilon^2$ terms gives $R(\eta,\xi)\approx[1+2\eta\varepsilon+(2\eta^2-1)\varepsilon^2]\exp(-\eta^2)$. The second term in square brackets is responsible for oscillations of the soliton's center, while the third term affects net gain experienced by the beam and leads to power oscillations with a period of $\pi/\Omega$. The averaging over one period of oscillations at $\Omega\gg1$ gives "effective" gain:

$$R_{\text{eff}}(\eta)=[1+(\eta^2-1/2)a^2]\exp(-\eta^2) \qquad (2)$$

which determines the parameters of the "guiding center" of the soliton. One can see that oscillations result in a decrease of the peak value of the effective gain and deformation of the entire gain landscape. It is well known that in a defocusing medium the shape of the static gain landscape determines the critical peak gain $p_i^{\text{cr}}$ above which dissipative solitons can form [14]. Since in a dynamical system the oscillations effectively reduce peak gain and deform the entire gain landscape experienced by the beam [see Eq. (2)], for sufficiently large amplitude $a$ of oscillations the peak gain level $p_i$ will become insufficient for soliton formation, even if the condition $p_i>p_i^{\text{cr}}$ was satisfied at $a=0$. In this case one expects decay of any input beam, i.e., disappearance of stable attractors in the system.

Figure 2(a) shows the dependencies of average soliton power and propagation constant [the instantaneous propagation constant value can be defined as $b=\lim_{d\xi\to0}[(1/id\xi)\ln(q|_{\xi+d\xi}/q|_\xi)]$] on the amplitude of oscillations $a$ obtained by direct integration of Eq. (1) for a $p_i$ value slightly exceeding the gain-guiding threshold $p_i^{\text{cr}}=1.63$ of the static Gaussian gain landscape. There exists a critical value of amplitude $a_{\text{cr}}$ beyond which solitons do not form. The dependence $U(a)$ is nearly parabolic, as suggested by the form of effective gain (2). In order to stress the validity of the effective gain approximation, we also show in Fig. 2(a) the dependence $U(a)$ obtained after replacement of the dynamically varying gain in Eq. (1) with the static "effective" gain profile from Eq. (2). Note that for larger $p_i$ values, when the amplitude of the oscillations of gain required for soliton destruction exceeds unity, the dependence $U(a)$ notably differs from the parabolic one [Fig. 2(b)], and the effective gain approximation accurately describes only the part of this dependence at $a\ll1$.

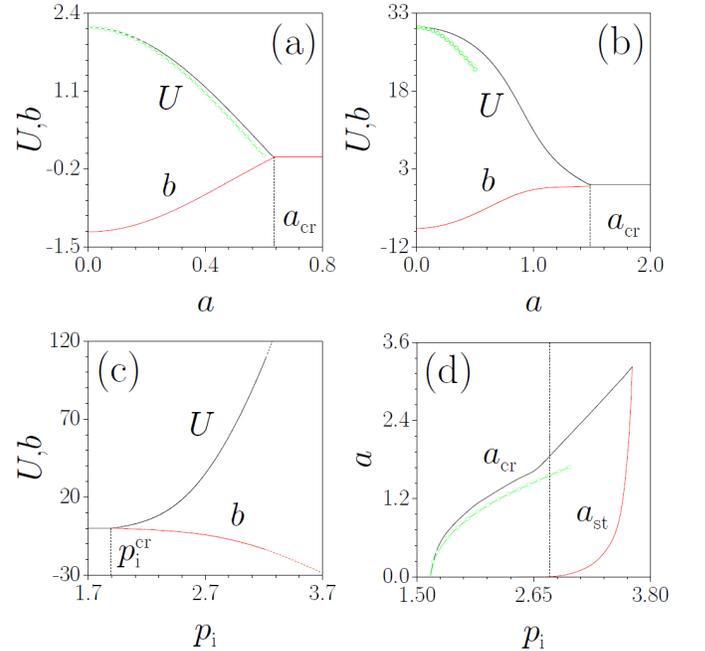

Fig. 2. Soliton energy flow $U$ and propagation constant $b$ versus amplitude of gain oscillations $a$ at $\Omega=2\pi$ for gain strength $p_i=1.8$ (a) and $p_i=2.5$ (b). $U(a)$ dependence obtained with effective gain model is shown by line with circles. (c) $U(p_i)$ and $b(p_i)$ dependencies at $\Omega=2\pi$, $a=0.8$. Solid lines - stable branches, dashed lines - unstable ones. (d) The domain of existence and stability on the plane $(p_i,a)$ at $\Omega=\pi$. Line with circles shows $a_{\text{cr}}$ obtained with effective gain model. Dashed line shows the border of the stability domain in the static gain landscape.

Note that Fig. 2 confirms that by changing the amplitude of the gain oscillations one can effectively manage the strength of the gain and soliton parameters. The presence of the threshold value of gain is also visible in Fig. 2(c), where we show $U(p_i)$ and $b(p_i)$ dependencies for fixed amplitude of oscillations. It should be mentioned that the increase of gain strength $p_i$ may result in desta-bilization of periodically oscillating states and the appearance of irregular oscillations [see dashed branches in Fig. 2(c)]. The domain of existence of the stable oscillating solitons on the plane $(p_i, a)$ is shown in Fig. 2(d). Any input beam decays at $a > a_{cr}$, while periodic oscillations occur for $a < a_{cr}$. The critical amplitude of oscillations vanishes when $p_i$ approaches the value at which gain-guiding occurs in static landscapes. Around this point the effective gain model gives an accurate prediction for critical oscillation amplitude (see the line with circles). While in the static gain landscape soliton destabilization occurs for $p_i > 2.8$ [dashed line in Fig. 2(d)], the oscillations of gain allow the stability domain to be extended considerably. In Fig. 2(d) solitons are stable above $a = a_{st}$ curve. Stable states disappear when $a_{cr}$ and $a_{st}$ curves join.

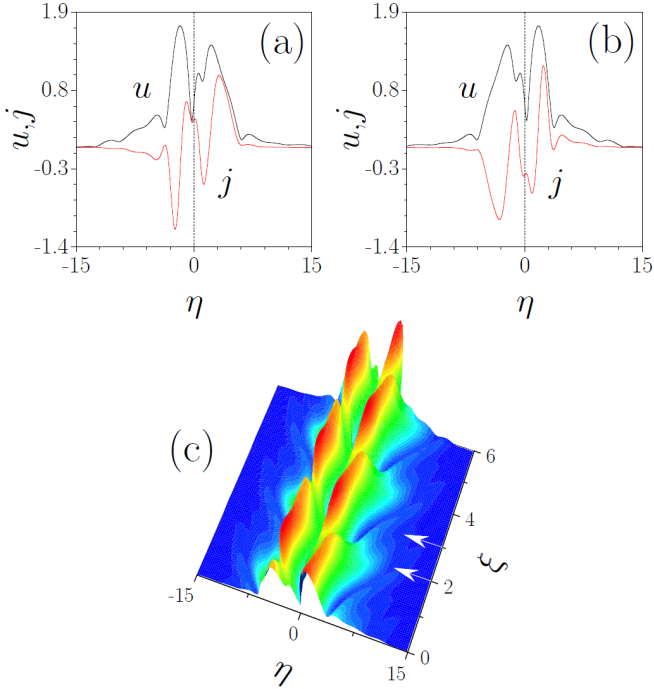

Fig. 3. The same as in Fig. 1, but for a two-channel gain landscape. The current in (a),(b) is divided by 150 for illustrative purposes.

It is worth noticing that very similar results were obtained for two synchronously oscillating gain channels separated by a distance $\delta\eta = 4$ (Fig. 3). They support collective modes with a dipole-like structure. The region of stability for such solitons is narrower than that for modes supported by a single amplifying channel.

Periodic oscillations of gain are not the only type of accelerated motion that can lead to modifications in effective gain. We found that the rotation of the two-dimensional gain channel $R = \exp[-(\eta - r\cos(\nu\xi))^2 - (\zeta - r\sin(\nu\xi))^2]$ with the frequency $\nu$ along the circle of radius $r = 4$ in the transverse plane leads to a similar effect. However, in rotating gain landscapes the amplitude, propagation constant, and energy flow of solitons do not change with distance once a steady-state regime is achieved, and only the center of the soliton moves along the circular trajectory. Such solitons leave behind them a relatively long asymmetric trace that is most pronounced for large rotation frequencies [see Fig. 4(a)], but close inspection of soliton tails reveals their exponential localization. Notice that several rotating gain channels equidistantly placed on a ring may support steadily rotating complexes of bright spots. One can distinguish several intensity minima placed between such bright spots, indicating the presence of several phase dislocations [Figs. 4(b) and 4(c)]. With the increase of the rotation frequency, the effective gain experienced by the soliton, and hence its power, rapidly drop down and at one point stable attractors disappear [Fig. 5(a)]. Notice that $U(\nu)$ dependence may be nonmonotonic for multi-channel gain landscapes. The critical rotation frequency $\nu_{cr}$ nearly coincides for solitons supported by one, two, and three amplifying channels. It monotonically increases with $p_i$. When $p_i$ approaches the value $p_i^{cr} \approx 2.65$ at which gain guiding occurs in static two-dimensional gain landscapes, the critical velocity vanishes [Fig. 5(b)].

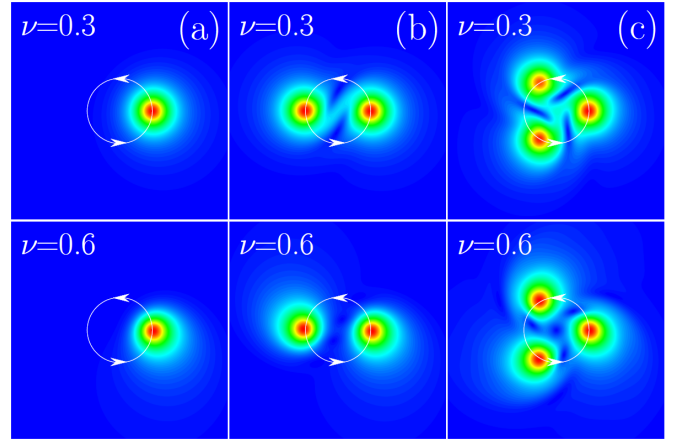

Fig. 4. Field modulus distributions for solitons supported by the rotating gain landscapes with one (a), two (b), and three (c) amplifying channels. Arrows show the rotation direction of the amplifying channels. In all cases $p_i = 2.8$.

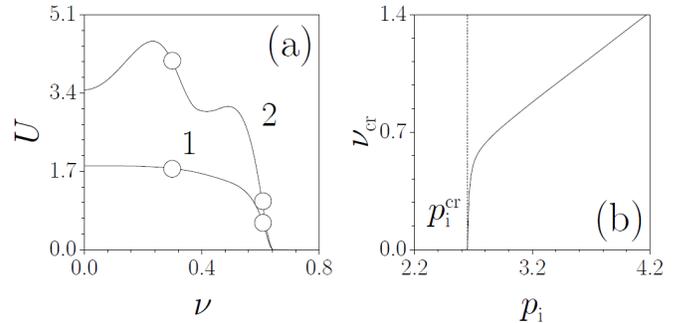

Fig. 5. (a) Energy flow versus rotation frequency in one- and two-channel gain landscapes at $p_i = 2.8$. Circles correspond to solitons shown in Fig. 4. (b) Critical rotation frequency versus gain parameter for one-channel gain landscape. Dashed line indicates gain-guiding point for non-rotating gain.

Summarizing, we predicted that stable attractors may form in dissipative systems with localized gain and defocusing nonlinearity, even if localized gain landscapes exhibit rapid and large-amplitude transverse oscillations. The accelerated motion of the gain landscape in the transverse plane allows the power and shapes of dissipative solitons to be controlled.